\documentclass[a4paper]{raa}           
\usepackage{graphicx,times}
\usepackage{natbib}
\usepackage{amssymb,amsmath}
\usepackage{rotating}
\bibpunct{(}{)}{;}{a}{}{,}

\usepackage[a4paper=true,pdftex=true,pagebackref=true]{hyperref}
\hypersetup{pdftitle = The title of my PDF, pdfauthor = My name, pdfsubject= The subject, pdfkeywords = keyword1 keyword2 keyword3} 
\hypersetup{colorlinks = true, linkcolor = green, anchorcolor = red, citecolor = blue, filecolor = red, pagecolor = red, urlcolor = red}

\begin{document}

   \title{A compilation of known QSOs for the Gaia mission
$^*$
\footnotetext{\small $*$ Supported by the National Natural Science Foundation of China.}
}

   \setcounter{page}{1}

   \author{Shi-Long Liao\inst{1,2} 
\and
Zhao-xiang Qi\inst{1,2}
\and
Su-fen Guo\inst{1,2}\\
\and
\and Zi-huang Cao\inst{3}
}

   \institute{Shanghai Astronomical Observatory, Chinese Academy of Sciences, Shanghai 200030, 
China; {\it shilongliao@shao.ac.cn}\\
        \and
             School of Astronomy and Space Science, University of Chinese Academy of Sciences, Beijing 100049, China\\
\and National Astronomical Observatory, Chinese Academy of Sciences, 20A Datun Road, Chaoyang District, Beijing, China \\
\vs \no
   {\small Received 2018 month day; accepted 2018 month  day}
}

\abstract{Quasars are essential for astrometric in the sense that they are spatial stationary because of their large distance from the Sun. The European Space Agency (ESA) space astrometric satellite Gaia is scanning the whole sky with unprecedented accuracy up to a few $\mu as$ level. However, Gaia's two fields of view observations strategy may introduce a parallax bias in the Gaia catalog. Since it presents no significant parallax, quasar is perfect nature object to detect such bias. More importantly, quasars can be used to construct a Celestial Reference Frame in the optical wavelengths in Gaia mission. In this paper, we compile the most reliable quasars existing in literatures. The final compilation (designated as Known Quasars Catalog for Gaia mission, KQCG) contains 1843850 objects, among of them, 797632 objects are found in Gaia DR1 after cross-identifications. This catalog will be very useful in Gaia mission.
}

   \authorrunning{S.-L. Liao et al. }            
   \titlerunning{A compilation of known QSOs for  Gaia mission}  
   \maketitle

%
\section{Introduction}           
\label{sect:intro}
The European Space Agency's (ESA's) Gaia mission, which launched in 2013, is a space-based, astrometric, photometric, and radial velocity all-sky survey at optical wavelengths (\citealt{Collaboration2016The}). As the successor to the Hipparcos mission (\citealt{perryman1997hipparcos}), Gaia will observe all objects with G magnitude down to ≃ 20.7 mag during its five-year mission. The main goal of this mission is to make the largest, most precise three-dimensional map of the Milky Way (\citealt{ESA-Gaia}). Compared to the Hipparcos mission, Gaia will bring a factor about 50 to 100 better in position accuracy (up to a few $\mu as$ level) and of a factor of about 10,000 more in star number (about 1 billion) (\citealt{eyer2011hipparcos}).  

Quasars (quasi-stellar objects or QSOs) are extremely distant and small in apparent size. They are ideal objects in establishing reference frame since they present no significant parallax or proper motion. The International Celestial Reference Frame (ICRF), which is the realization of the International Celestial Reference System (ICRS) at radio wavelengths (\citealt{Arias1995The}), consists of 3414 such compact radio objects in its second realization (\citealt{Boboltz2015The}). Gaia will observe about 500,000 QSOs (\citealt{Andrei2007A}), which will allow us to build a rotation-free Celestial Reference Frame in the optical wavelengths and meet the ICRS specifications. Besides, the principle that Gaia uses two fields of view  to observe might lead a global parallax bias in the Gaia catalog (\citealt{van2005rights};\citealt{ButkevichZero}; \citealt{butkevich2017impact},\citealt{Liao2017The}).  Since the QSOs' parallaxes can be treated as zero, they are idea objects to detect this parallax bias. 

Over the past few decades, the surveys such as 2dF/2QZ QSO's survey (\citealt{Croom2004The}), the Solan Digital Sky Survey (SDSS) (\citealt{Fukugita1996The}) contribute the majority of the QSOs found in optical wavelength. There were thousands of compact extragalactic sources observed by the radio VLBI technique and are listed in the ICRF2 (\citealt{Boboltz2015The}), VLBA (\citealt{Beasley2002The}; \citealt{Fomalont2003The}), VLA (\citealt{VLA_m}), and JVAS catalogs (\citealt{Patnaik1992Interferometer}; \citealt{Browne1997Interferometer};\citealt{Wilkinson1998Interferometer}). And there are also QSOs discovered in X-ray wavelength, such as Swift X-ray Point Source catalog (\citealt{Evans20131SXPS}). Three decades ago, Véron-Cetty and Véron gathered all those quasars into a single catalog (hereafter V$\&$V), which has been updated since then until 2010 (\citealt{V1984,V2006A,V2010A}). The latest version of their catalog included 133 336 quasars and 34 231 active galaxies. J.Souchay constructed three successive versions of a compilation of quasars catalogue designated the Large Quasar Astrometric Catalogue (LQAC) with the aim to give useful astrometric data concerning all quasars (\citealt{Souchay2009The,Souchay2012The, Souchay2015The,Souchay2017The}). The latest version (called LQAC3), which contains 321 957 objects, was published in early 2015. With the new data release of SDSS (\citealt{Abolfathi2017The}, \citealt{P2017The}, see \url{http://www.sdss.org/dr14/} for more detail), the surveys such as the Large Sky Area Multi-Object Fiber Spactroscopic Telescope (LAMOST) (\citealt{Zhao2012Large}; \citealt{Cui2012LAmost}) and the Wide-field Infrared Survey Explorer (WISE) (\citealt{Wright2012THE};\citealt{Secrest2015Identification}), the number of quasars discovered in recent years increase rapidly. The Million Quasars (MILLIQUAS) Catalog (\citealt{Flesch2017VizieR}; \citealt{Flesch2015The}) compiles all the known or candidate AGNs/QSOs through 2017 August,  which contains 1,998,464 objects including all-sky radio/X-ray associated objects. 

A Gaia Initial QSO Catalog (GIQC) was compiled with the LQAC-2 list, the QSOs from SDSS DR10 and the BOSS selection,  the VLBI QSOs (\citealt{Andrei2014The}) in 2014. However, several reasons lead us to implement a new compilation of the QSOs catalog specific for the Gaia mission. First, the spatial uniformity of GIQC is needed to improve, see Figure \ref{LQAC3_Source} for the sky distribution of LQAC3. Second, to build up the Gaia reference frame and to detect the parallax bias of the Gaia catalog, it is crucial to maximize the number of QSOs in optical wavelength. However, none of the current QSO catalogs are updated to present date to include all the new discovered QSOs:  V$\&$V contains the new QSO discovered until 2010; LQAC was only updated to early 2015, and doesn't contain the QSO objects from SDSS DR14, WISE and LAMOST DR5 (see \url{http://dr5.lamost.org} for more detail); While MILLIQUAS missed the QSOs from LAMOST DR5; Third, false identification of QSOs may lead to a wrong detection of the parallax bias. The catalog such as MILLIQUAS contained a large part of QSO candidates, which are not suitable for the Gaia mission.  

This paper is organized as follows: Section two is the introduction of the data used; Section three is the compilation of the QSOs, and the last section is the discussion.

\section{Data used}
The purpose of this compilation is to provide positions of known QSOs, which can be used to cross-match with the Gaia observations. To maximize the number of QSOs/AGNS, we choose four samples for their huge number of reliable QSOs/AGNs. We started from the LQAC3 QSOs list. LQAC3 contains all the known QSOs discovered in 2dF/2QZ survey, the DR10Q of the SDSS-III (\citealt{Bolton2012Spectral}; \citealt{Paris2012The}) and the VLBI QSOs listed in ICRF2, VLBA, VLA, and JVAS catalogs. Here in this compilation, the QSOs from SDSS in LQAC3 will be updated with the latest spectroscopically confirmed QSOs in the SDSS-DR14 Quasar Catalog (DR14Q) (\citealt{P2017The}) by using data from the Baryon Oscillation Spectroscopic Survey (BOSS; \citealt{Eisenstein2015SDSS}, \citealt{Dawson2012The}) and photometrically selected QSO in SDSS.  And then enter the spectroscopically confirmed QSOs in LAMOST DR5 using LAMOST spectroscopic data. At last we complement the mid-IR color selection AGNs by using the WISE data (\citealt{Wright2012THE}; \citealt{Secrest2015Identification}). 

\subsection{The Large Quasar Astrometric Catalog}
\begin{figure}
   \centering
   \includegraphics[width=14.0cm, angle=0]{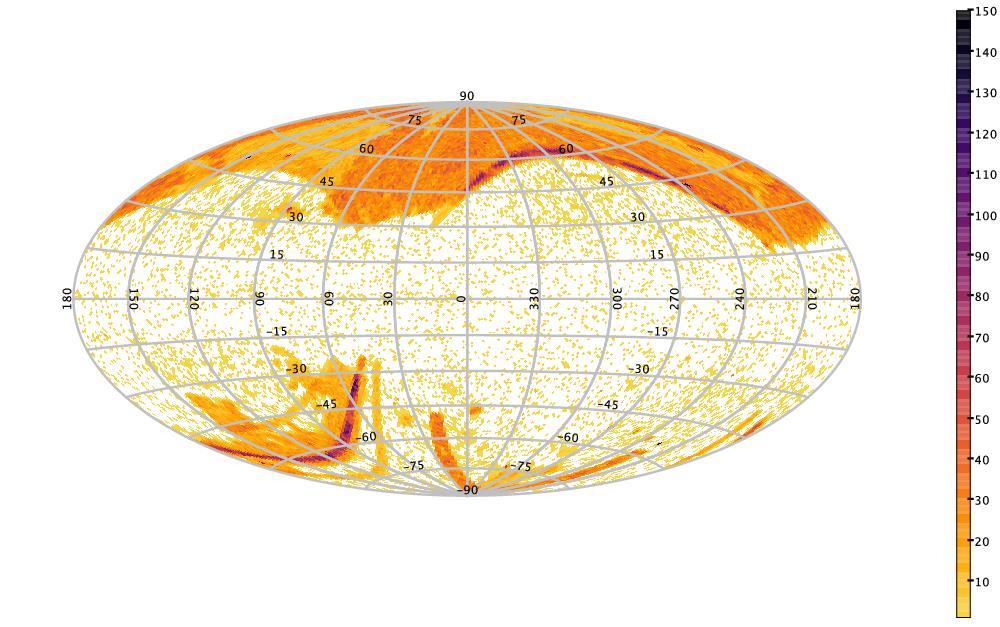}
   \caption{The density distribution plot of the LQAC3 sample of AGN/QSO sources with Aitoff projection in Galactic coordinates. The shader on the right is the sky density of objects (number per square degrees,the same applies hereinafter). } 
   \label{LQAC3_Source}
   \end{figure}

The LQAC3 catalog is the latest version of the three successive versions of a compilation of known quasars catalogue, which contains 321 957 objects including 14128 AGNs and 1183 BLLac. It was built by carrying out the cross-identification between the existing catalogs of quasars chosen for their huge number of objects up to 2015. LQAC3 includes all the available data related to magnitudes, radio fluxes, and redshifts. The sky density distribution of LQAC3 is shown in Figure \ref{LQAC3_Source}, and the composition can be found in Table \ref{tab1}.

\begin{table}
\centering
\bc
\caption[]{Number of quasars from each catalog \label{tab1} in LQAC3 compilation.}
\setlength{\tabcolsep}{1pt}
\small
 \begin{tabular}{cccc}
  \hline\noalign{\smallskip}
Origin & Flag & Nature &Number\\
  \hline\noalign{\smallskip}
ICRF2&A&radio&3414\\
VLBA&B&radio&7213\\
VLA&C&radio&1858\\
JVAS&D&radio&2118\\
SDSS&E&optical&262535\\
2QZ&F&optical&23660\\
2df-SDSS LRG &G&opticl&9058\\
FIRST&H&radio&969\\
HB &I&optical and radio&6720\\
V$\&$V&M&optical and radio&79692\\
  \noalign{\smallskip}\hline
\end{tabular}
\ec
\tablecomments{0.86\textwidth}{This table is taken from LQAC3 paper (\citealt{Souchay2015The}). The HB catalog is the abbreviation of Hewitt $\&$ Burbridge catalog (\citealt{Hewitt1993A}).}
\end{table}

\subsection{The SDSS QSO catalog}
\begin{figure}
   \centering
   \includegraphics[width=14.0cm, angle=0]{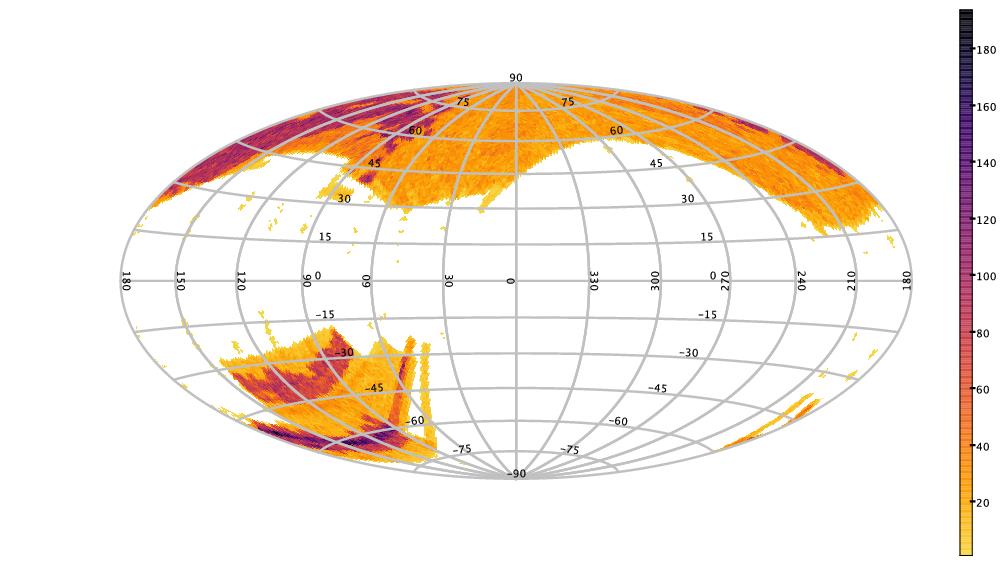}
   \caption{The density distribution plot of the SDSS DR14Q sample of QSO sources with Aitoff projection in Galactic coordinates.} 
   \label{SDSS_Source}
   \end{figure}

The Sloan Digital Sky Survey (SDSS) is a large imaging and spectroscopic survey, which uses the Sloan Foundation 2.5m optical telescope at Apache Point Observatory in New Mexico for the Northern sky survey, and the 2.5m Du Pont optical telescope at Las Campanas Observatory in Chile after being extended to the Southern Sky. It has progressed through four phases including SDSS-I (2000-2005), SDSS-II (2005-2008), SDSS-III (2008-2014), and SDSS-IV (2014-2020). The SDSS-DR14 Quasar Catalog (DR14Q) used here comes from the second data release of the extended Baryon Oscillation Spectroscopic Survey (eBOSS) (\citealt{Eisenstein2015SDSS}, \citealt{Dawson2012The}) of the  SDSS-IV. This catalog includes all SDSS-IV/eBOSS objects that were spectroscopically targeted as quasar candidates and that are confirmed as quasars (\citealt{P2017The}). The SDSS DR14Q quasar catalog also contains all the quasars observed as part of SDSS-I/II/III. The total number of the QSOs in SDSS DR14Q are 526 356. For each object, the catalog presents five-band (u, g, r, i, z) CCD-based photometry. The spatial distribution is largely homogeneous near the North Galactic Cap and the South Galactic Cap. See Figure \ref{SDSS_Source} for the spatial density distribution. 

\subsection{The QSOs in LAMOST DR5}
\begin{figure}
   \centering
   \includegraphics[width=14.0cm, angle=0]{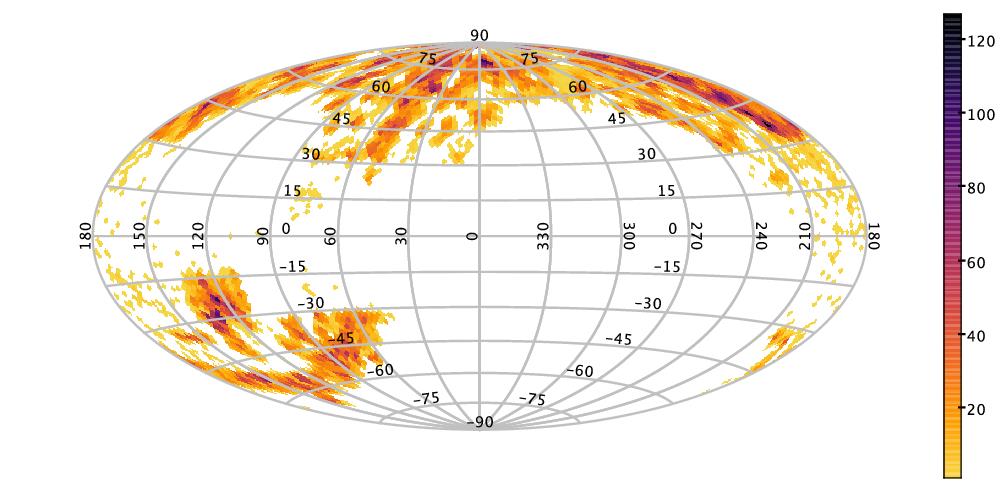}
   \caption{The density distribution plot of AGN/QSO sources in LAMOST with Aitoff projection in Galactic coordinates. } 
   \label{LAMOST_Source}
   \end{figure}

The Large Sky Area Multi-Object Fiber Spectroscopic Telescope (LAMOST) is a reflecting Schmidt telescope with a wide field of view of 5 degrees in diameter and an effective aperture of about 4 m (\citealt{Zhao2012Large}; \citealt{Cui2012LAmost}), which can record 4000 celestial object spectra simultaneously with its 4000 fibers. The telescope began its first spectroscopic regular survey in September 2012. The fifth data release (DR5) has published online in late 2017, which include 51,133 quasars in the LAMOST general catalog (see \url{http://dr5.lamost.org} for more detail). Due to the lack of systematic spectroscopic survey, the number of spectroscopically confirmed quasars remains very small in low Galactic latitude. With the help of the LAMOST Spectroscopic Survey of the Galactic Anti-center (LSS-GAC) (\citealt{Liu2014LSS}; \citealt{Yuan2014LAMOST}), 151 unique quasars from the LAMOST in the anti-center of Galactic are discovered (\citealt{Huo2017Quasars}). These quasars from LAMOST will play an important role in the proper motion and parallax validation of Gaia results in the highly dust extinct Galactic disk regions. See Figure \ref{LAMOST_Source} for the sky density distribution. 

\subsection{The WISE AGN catalog}
\begin{figure}
   \centering
   \includegraphics[width=14.0cm, angle=0]{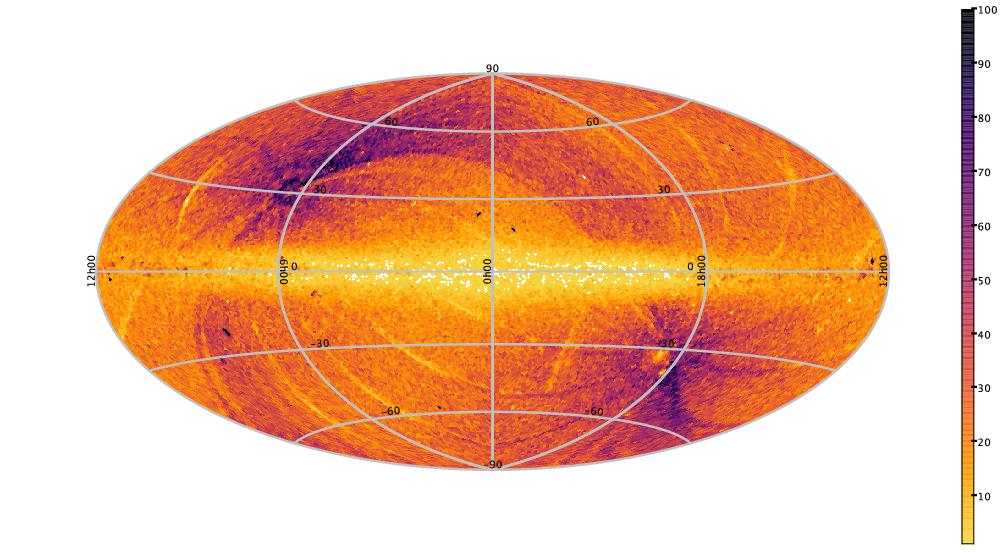}
   \caption{  Density plot, Aitoff projection in Galactic coordinates, WISE sample of AGN/QSO sources. As indicated by the author, the over-density of sources at the ecliptic poles is due to deeper WISE coverage; While the under- abundance along the Galactic plane is due to source confusion (\citealt{Secrest2015Identification}).}
   \label{ALL_WISE_Source}
   \end{figure}

The Wide-field Infrared Survey Explorer (\citealt{Wright2012THE}) is a satellite with a 40 cm aperture that launched by NASA in 2009 December, and begun its survey in 2010 January and finished the first all-sky cryogenic survey in 2010 August. After that, an additional $30\%$ of the sky was mapped from 2010 September to 2011 February. The satellite scanned the whole sky at mid-IR at 3.4, 4.6, 12 and 22 $\mu m$ (W1, W2, W3, and W4, respectively). Sources in WISE catalog are classified as active galactic nuclei (AGNs) from a two-colour infrared photometric criterion with the observations of WISE data (\citealt{Secrest2015Identification}). This catalog (mid-IR AGNs, abbreviated as MIRAGNs) contains the all-sky sample of 1.4 million sources. As estimated by the author, the probability of a star detected in this optical survey will be misidentified as a QSO is smaller than $0.041\% $ per source, which makes MIRAGNs highly promising for celestial reference frame work for its huge number of uniformly spatial distributed compact extragalactic sources. The sky density distribution of  MIRAGNs is shown in Figure \ref{ALL_WISE_Source}.

\section{The compilation}
\subsection{Cross-identifications between catalogs}
\begin{figure}
   \centering
   \includegraphics[width=14.0cm, angle=0]{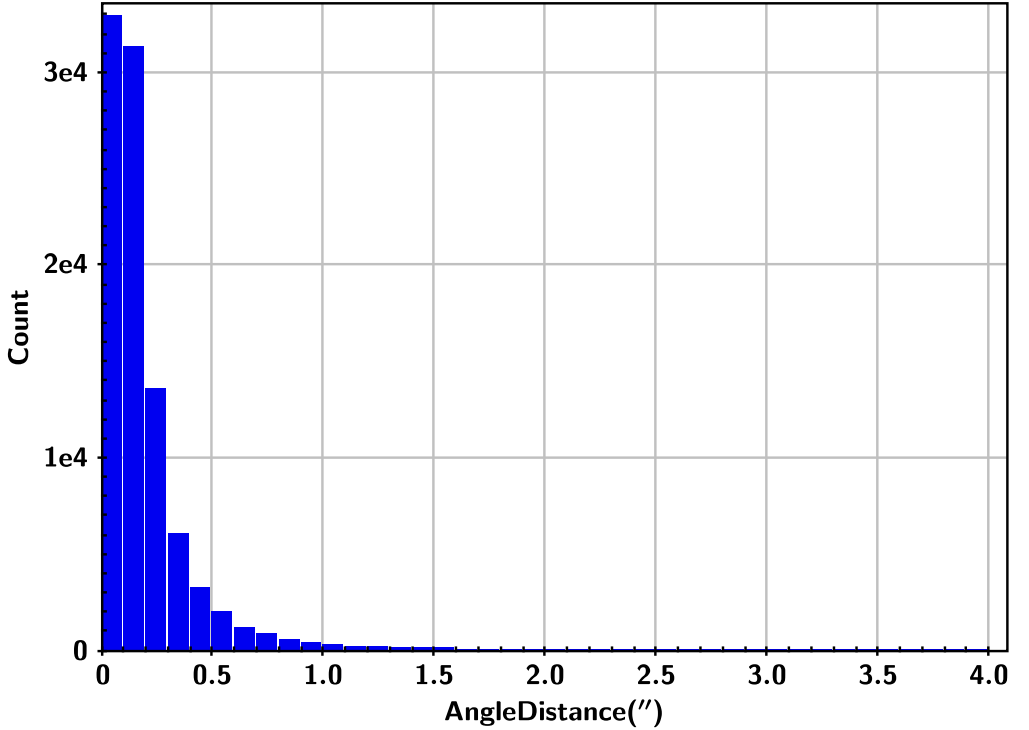}
   \caption{Histogram of angle distance of the cross-identified quasars between LQAC3 and MIRAGNs sample. The abscissa is the angle distance between two identified targets, and the ordinate is the number of matching.} 
   \label{angDistance}
   \end{figure}

The criteria of being identified as the same object in the cross-identifications is the angle distance between two objects smaller than a certain value. As shown in Figure \ref{angDistance}, the majority of the angle distances between two objects are smaller than 2 arcsecond. The larger the angle distance, the higher possibility of false identifications. Thus, the search radius of the angle distance in the cross-identifications is set to 2 arcsecond. Even so, the risk of finding a false pair or missing identifications are unavoidable. In the first situation, the redshift information can be used as a second criteria as adopted in the construction of LQAC (\citealt{Souchay2009The}).  A significant difference in redshift value indicates that they are a pair of coincident quasars. In contrast, close values of redshifts with a small angular separation suggest that these two objects are a false identification pair, which will be discarded.

\subsection{The final catalog}
\begin{figure}
   \centering
   \includegraphics[width=14.0cm, angle=0]{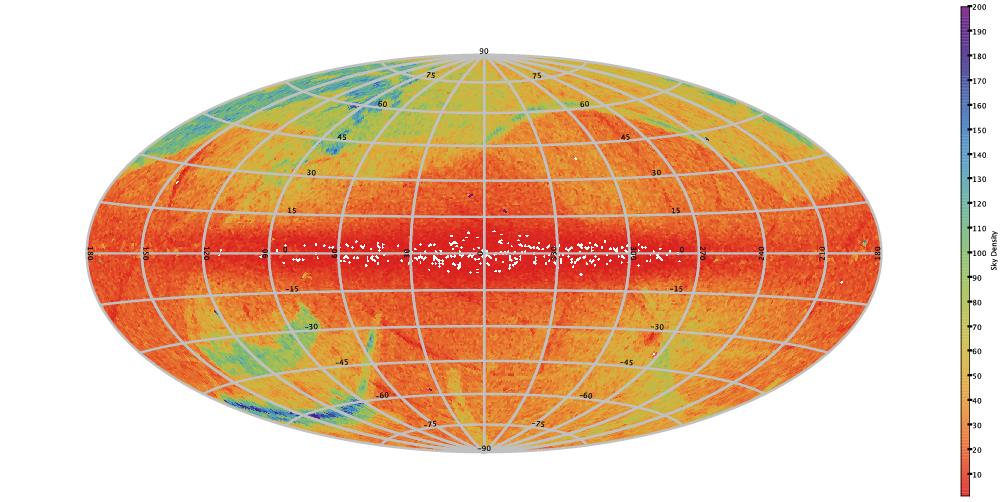}
   \caption{ The sky density distribution of the final QSO catalog. Aitoff projection in galactic coordinate.} 
   \label{Total_qso}
   \end{figure}

\begin{figure}
   \centering
  \includegraphics[width=6.8cm,angle=0]{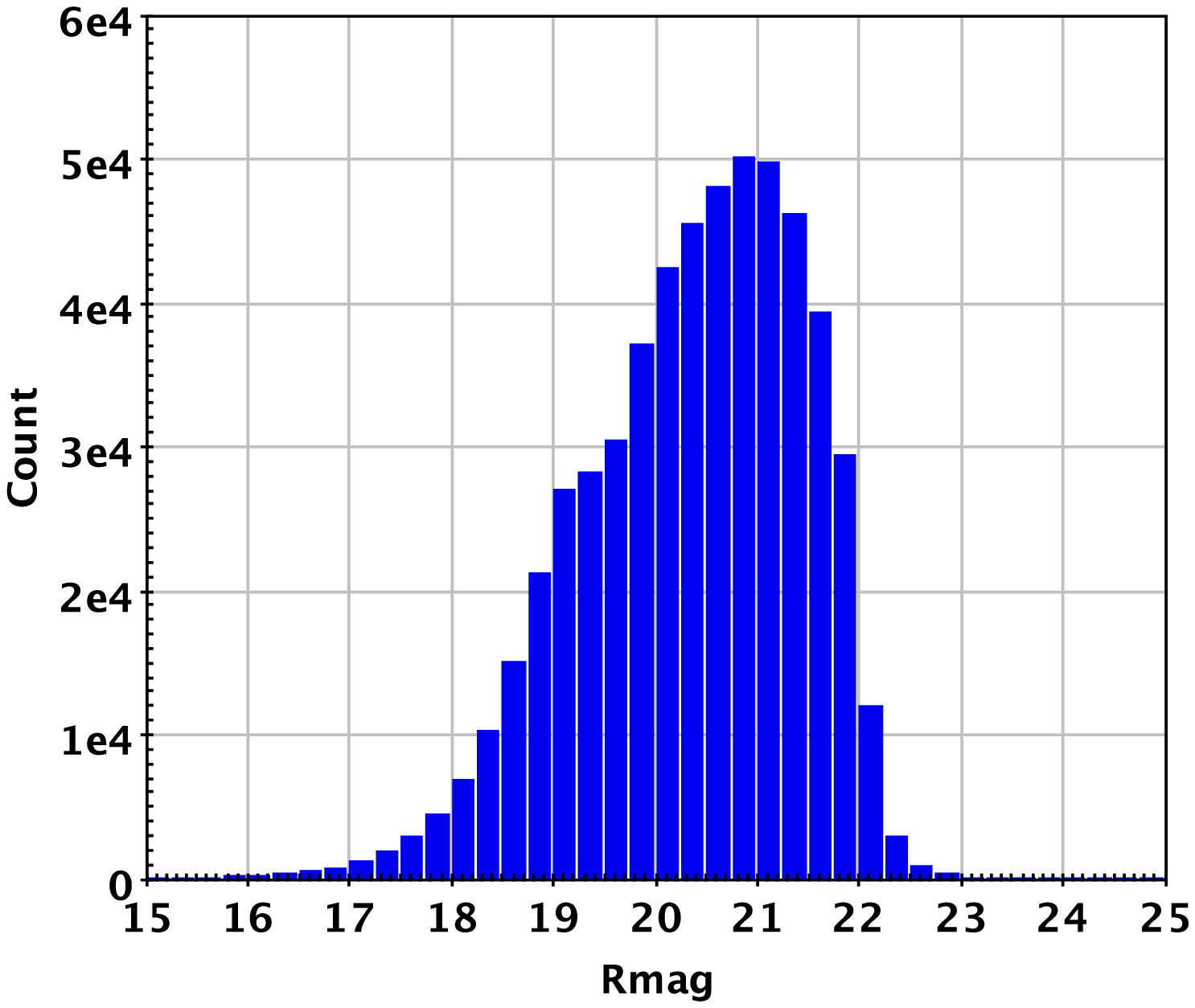}
   \includegraphics[width=7.1cm, angle=0]{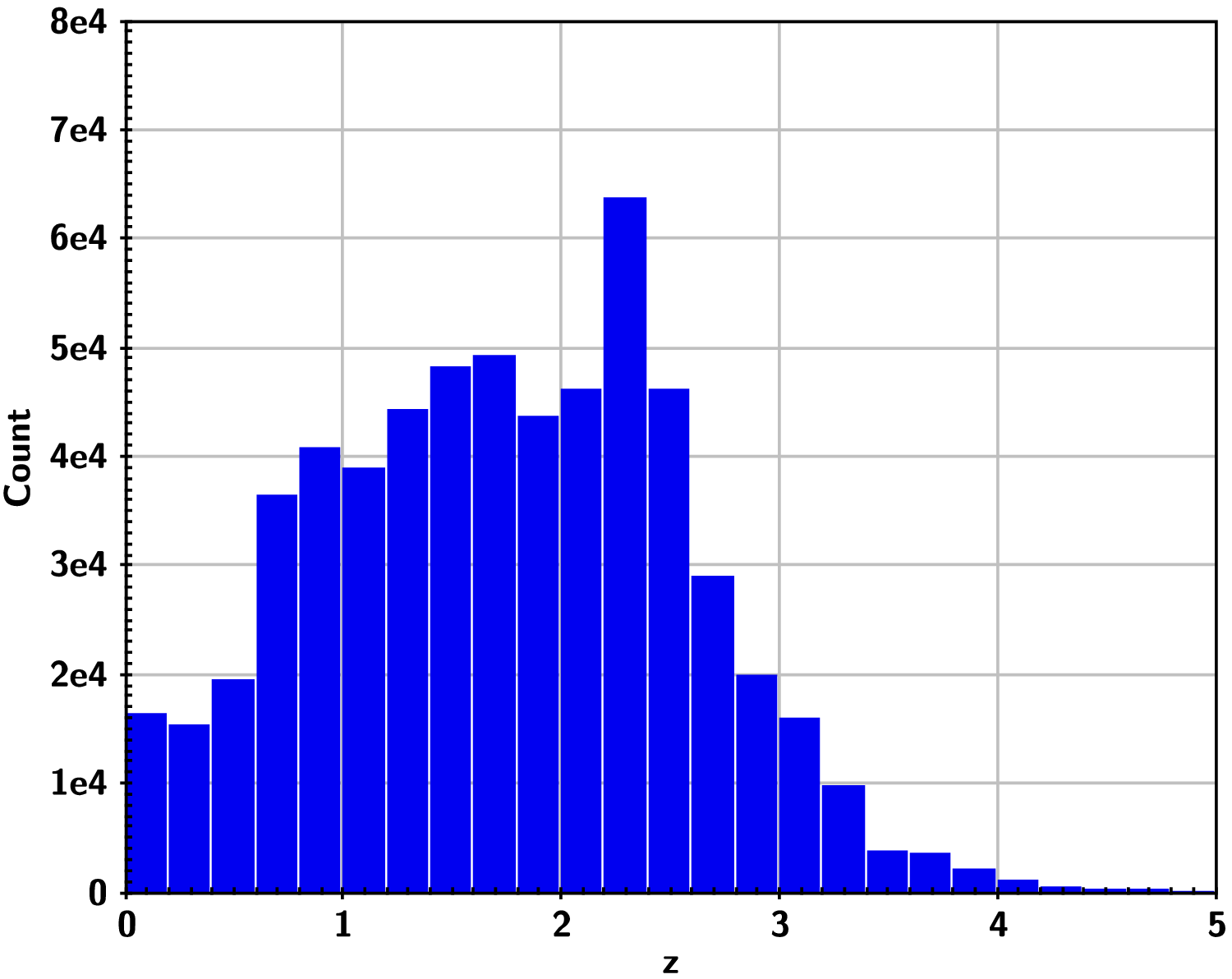}
   \caption{The histogram of magnitude in R band (left) and the red shift distribution of the final compilation (right) sample of AGN/QSO sources.} 
   \label{Fiall_Z}
   \end{figure}

 The final catalog (designated as Known Quasars Catalog for Gaia mission, KQCG) contains 1843850 QSOs in total after compilation. The spatial distribution of these objects are shown in Figure \ref{Total_qso}. The density is higher with the deeper coverage of the SDSS and LAMOST survey area near the North/South Galactic Cap. The average density of the sky distribution is about 45 objects per $deg^2$. The SDSS DRQ14 and MIRAGNs catalog contribute the majority of the optical QSOs in this compilation. See the Table \ref{tab2} for the composition of the final catalog.  The redshifts of the final catalog are shown in the right panel of Figure \ref{Fiall_Z}. One can see that the redshift of these QSOs are below $z=4$. The histogram of the magnitude in R band (the effective wavelength midpoint of R band is $\lambda_{eff} \approx 658 nm$) is shown in the left panel of Figure \ref{Fiall_Z}. 

\begin{figure}
   \centering
   \includegraphics[width=7.00cm, angle=0]{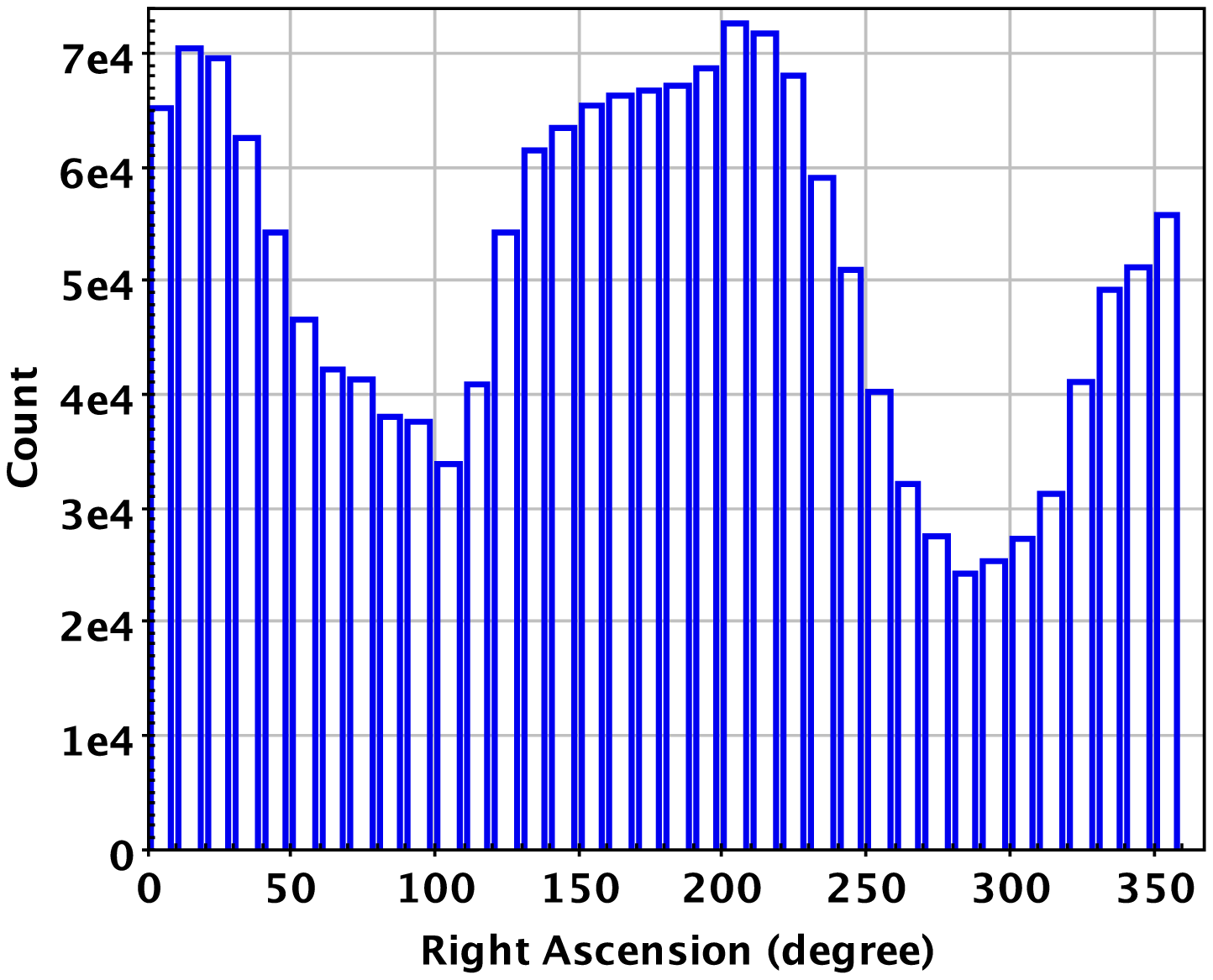}
   \includegraphics[width=7.0cm, angle=0]{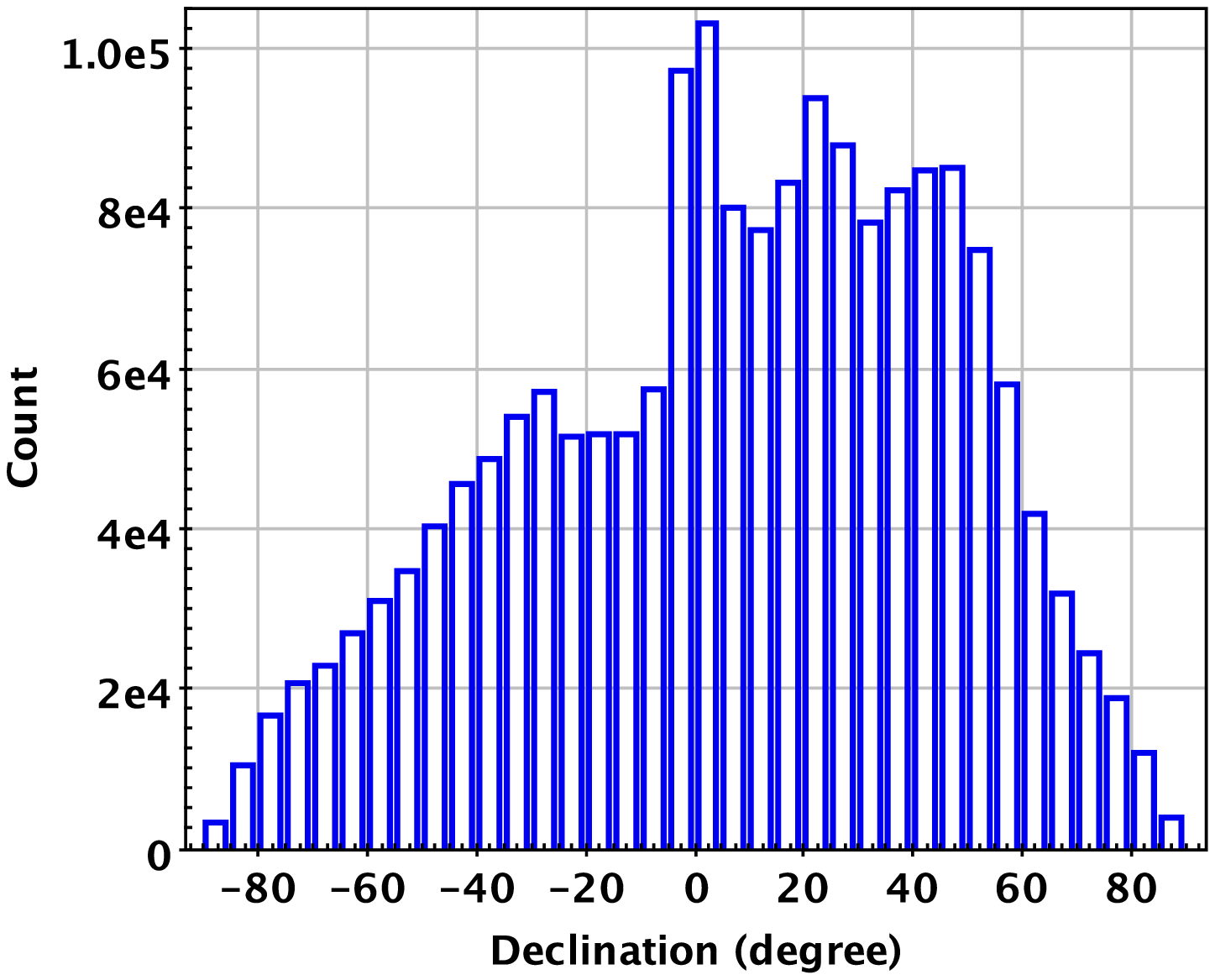}
   \caption{Histogram of the number of AGN/QSO sources of the final sample with respect to the right ascension (left) and  declination (right).} 
   \label{Equ_count}
   \end{figure}

\begin{figure}
   \centering
   \includegraphics[width=7.0cm, angle=0]{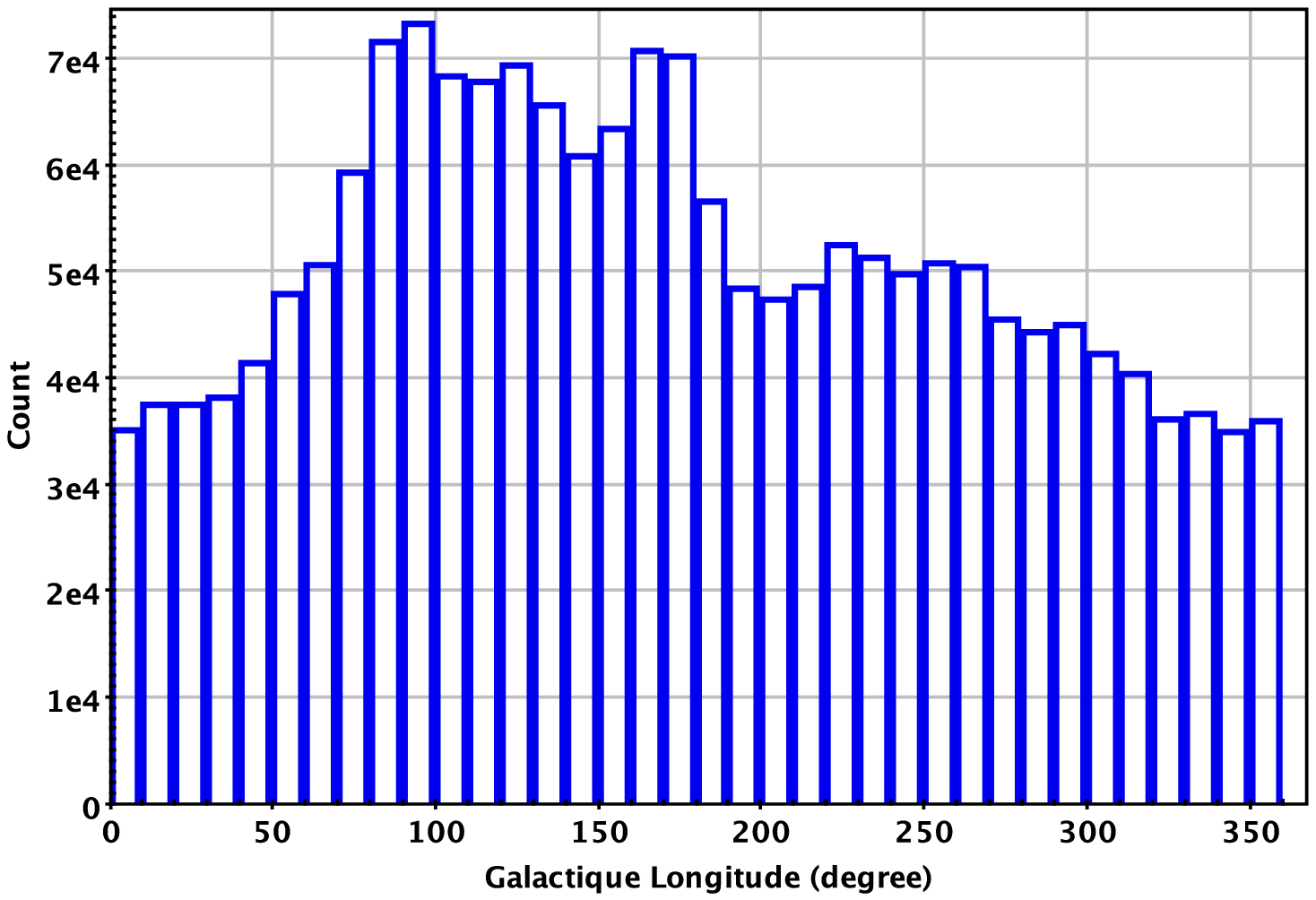}
   \includegraphics[width=7.00cm, angle=0]{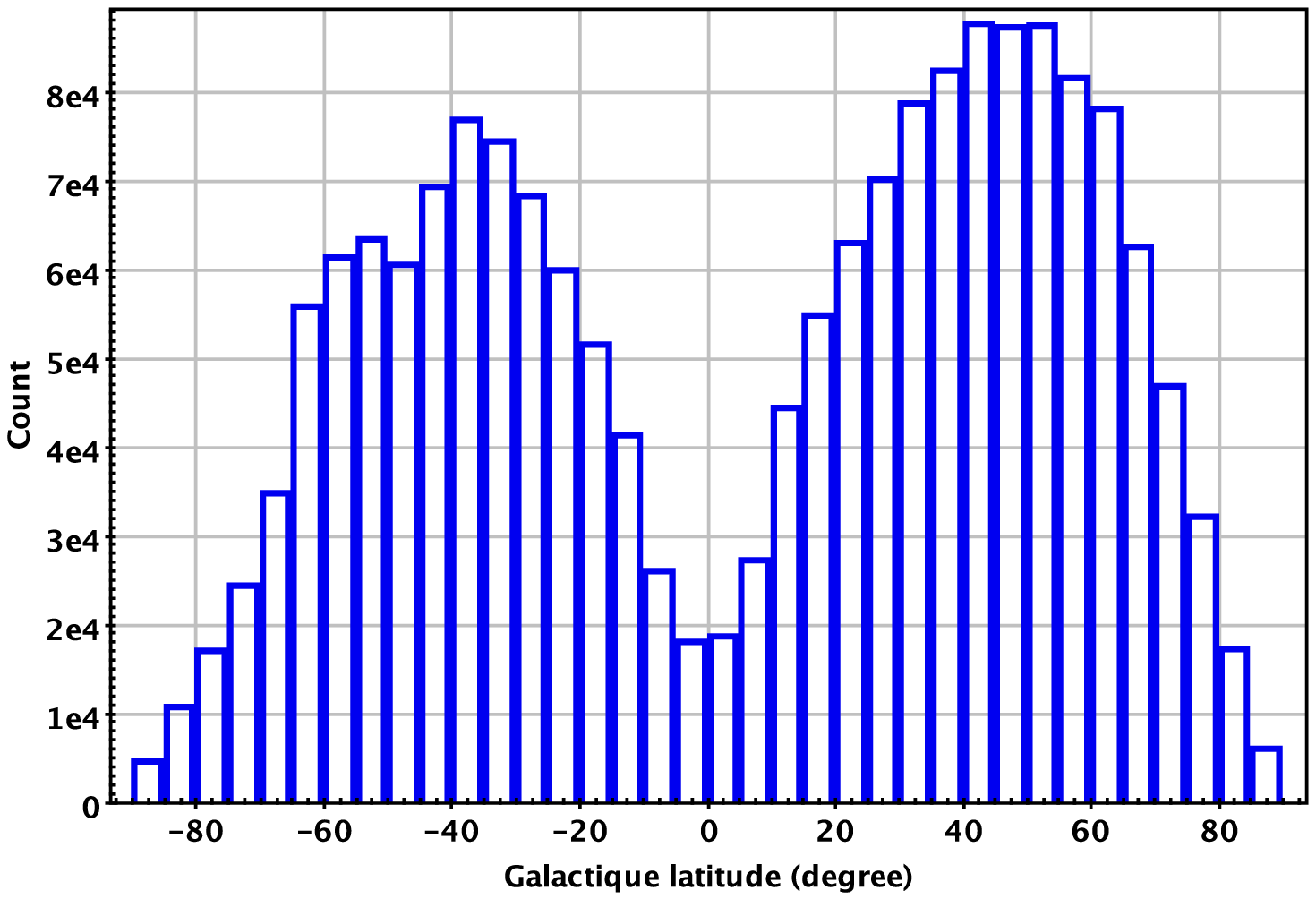}
   \caption{Histogram of the number of AGN/QSO sources of the final sample with respect to the Galactic longitude (left) and  Galactic latitude (right).} 
   \label{Gal_count}
   \end{figure}

Figure \ref{Equ_count} and Figure \ref{Gal_count} are the sky coverage of the final catalog with respect to the equatorial and galactic coordinates, respectively. In the left panel of Figure \ref{Equ_count}, the two depletion zones $ [70^{\circ}-120^{\circ}]$ and $ [260^{\circ}-310^{\circ}]$ are due to the vertical crossing of the Galactic plane. The detection is more sufficiency in northern hemisphere than the southern part in equatorial coordinates, which can be clearly visible in the right panel of Figure \ref{Equ_count}. The distribution of sources is rather homogenous with respect to the galactic longitude, and drops drastically in the neighborhood of the galactic plane and the galactic pole with respect to the galactic latitude, as shown in Figure \ref{Gal_count}.



\subsection{Description of the final catalog}
The final catalog contains 1843850 objects in total. A large part of them come from the WISE and SDSS survey. The purpose of this compilation is to provide the positions of the QSOs, in the following we present the items of this compilation. An example of this final catalog is shown in Table \ref{tab3}.

\begin{itemize}
    \item[-]Columns 1 gives the reference name of the quasar in the catalog. If there is a name  from its origin catalog, the original name is used. Otherwise, it will be designated by its equatorial coordinates calculated in degree and truncated to the arcsecond level.
    \item[-] Columns 2 and 3 are the equatorial coordinates ($\alpha, \delta$) of the object given by the original catalog.
\item[-] Column 4 gives the redshift value z of each quasar.
\item[-] Columns 5 to 9 give the apparent magnitudes in the u, g, r, i, z bands, respectively. These photometric values come from their origin catalogs. When the magnitude is not available, the value will be set empty.
\item[-]  Column 10 provides the reference catalog or flag indicating the presence of the quasar in one of the 12 catalogs from A to P as LQAC3 does, see Table \ref{tab2}.
\item[-] Columns 11-13 give the W1, W2 and W3 magnitude information of the sources from MIRAGNs.
\end{itemize}

\begin{table}
\centering
\bc
\caption[]{Number of quasars for each catalog \label{tab2} after compilation in KQCG.}
\setlength{\tabcolsep}{1pt}
\small
 \begin{tabular}{ccccc}
  \hline\noalign{\smallskip}
Origin & Ref/Flag & Nature &Number& Number found \\
&&&& in Gaia DR1\\
  \hline\noalign{\smallskip}
ICRF2&A&radio&3414&2372 \\
VLBA&B&radio&7213&4450\\
VLA&C&radio&1858&1269\\
JVAS&D&radio&2118&1406\\
SDSS&E&optical&499127&245982\\
2QZ&F&optical&23660&18203\\
2df-SDSS LRG &G&opticl&9058&2476\\
FIRST&H&radio&969&920\\
HB &I&optical and radio&6720&6064\\
V$\&$V&M&optical and radio&79692&7584\\
LAMOST&O&optical&24666&22252\\
MIRAGNs&P&optical&1260635&519412\\
  \noalign{\smallskip}\hline
\end{tabular}
\ec
\tablecomments{0.86\textwidth}{The reference/flag symbols are adopted from LQAC3, with the 'O' (LAMOST) and 'P' (MIRAGNs) are the two new symbols. The objects found in Gaia DR1 will be added with a flag "Q".}
\end{table}

%
%

\begin{figure}
   \centering
   \includegraphics[width=14.0cm, angle=0]{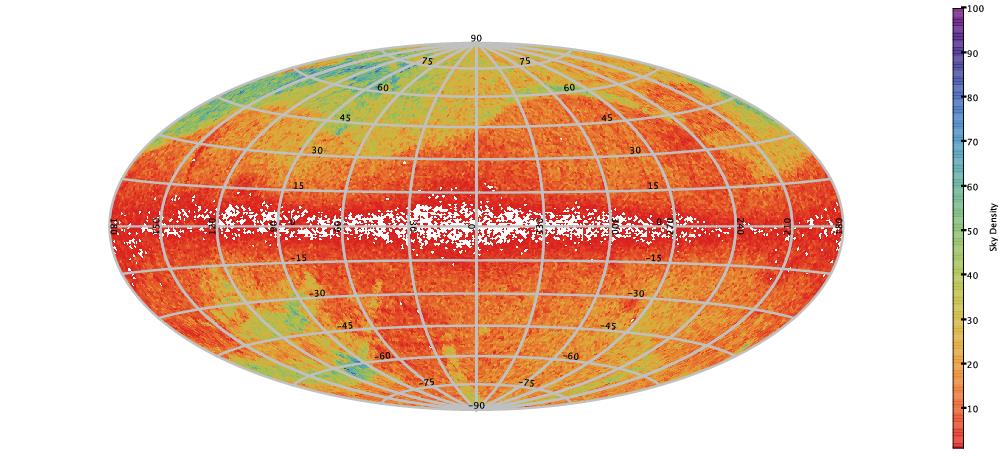}
   \caption{ The sky density distribution of the QSOs in KQCG found in Gaia DR1. Aitoff projection in galactic coordinate.} 
   \label{DR1_qso}
   \end{figure}

\begin{figure}
   \centering
   \includegraphics[width=14.0cm, angle=0]{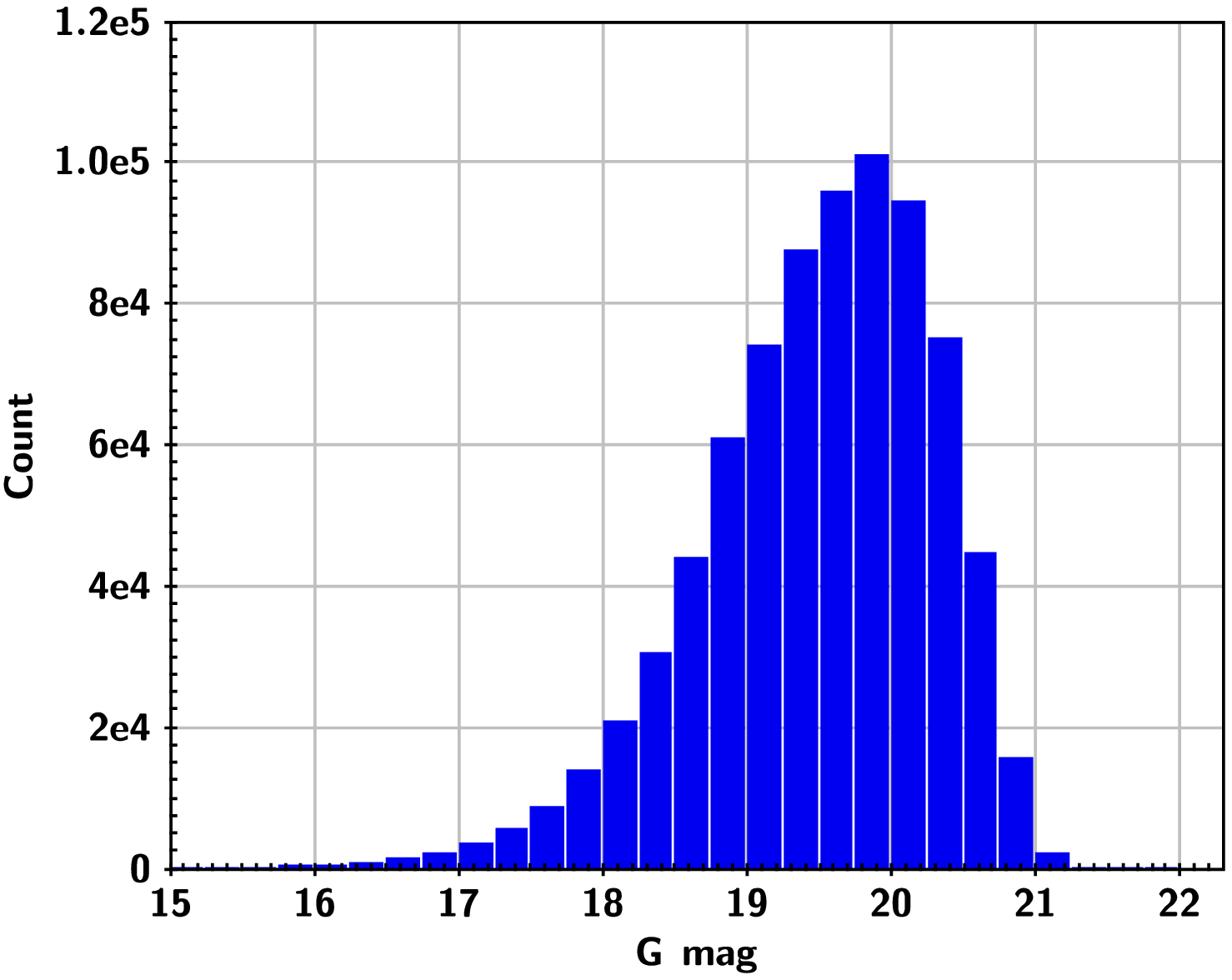}
   \caption{The histogram of Gaia G magnitude of AGN/QSO sources after cross-matching with Gaia DR1.} 
   \label{Final_X_DR1_Mag}
   \end{figure}

\section{Discussions}\label{sect:discussion}
We compiled the most reliable QSOs from LQAC3, SDSS DR14Q, LAMOST DR5 QSO catalog, and the MIRAGNs from WISE survey. The final catalog contains 1843850 objects in total. To evaluate the possible detected QSOs of this catalog in Gaia mission, we cross match this catalog with the Gaia DR1 catalog (\citealt{brown2016gaia}; \citealt{lindegren2016gaia}). The cross-match criteria is set to 2 arcsecond, and 797632 objects are found. Among of these objects, there are 5368 objects are marked as radio QSOs in LQAC3. The sky density distribution can be found in Figure \ref{DR1_qso}, the  histogram of their Gaia G magnitudes can be found in Figure \ref{Final_X_DR1_Mag}.  

This compilation of QSOs can be used in the Gaia mission in the following three aspects: 1) They can be selected to build the Gaia celestial reference frame. With such a large number of QSOs and uniform spatial distribution, one can statistics analysis the overall properties of this new Celestial Reference Frame with respect to their spatial distribution, accuracy, magnitude distribution. 2). As the parallaxes and the proper motions of the QSOs can be treated as zero, they can be used to detect the parallax bias of the Gaia catalog and as additional quality indicators to evaluate the overall quality of the Gaia catalog. 3). A comparison of those QSOs in radio wavelength with optical counterparts can be made between the Gaia solution and the VLBI solution. More importantly, these QSOs in radio wavelength with optical counterparts can be used to link the ICRF between the radio and optical band.

\begin{sidewaystable}[h]
\caption{A sample of few lines of \label{tab3} the final catalog.} 
\centering 
\begin{tabular}{{ccccccccccccc}} 
  \hline\noalign{\smallskip}
Name&RA$\_$Deg & DEC$\_$Deg&z&Umag&Gmag&Rmag&Imag&Zmag&Ref&W1-W2&W2-W3&W1$\_$Mag\\
  \hline\noalign{\smallskip}
LAMOST$\_$330.263-0.88341&330.2630600000&-0.8834100000&0.2121040000&16.85&16.64&16.21&15.78&15.98&O& & & \\			
LAMOST$\_$333.739-1.14075&333.7391700000&-1.1407500000	&0.0909574000&20.40&18.61	&17.53&17.01&16.58&O&	& & \\	
LAMOST$\_$332.883-1.13998&332.8827200000&-1.1399800000&1.0549400000&20.47&18.59	&17.56&17.15&16.83&O& 	& &\\	
LAMOST$\_$332.899-0.95608&332.8991100000&-0.9560800000&1.0639500000&19.83&18.49&17.90&17.48&17.28&O&&&\\			
LAMOST$\_$332.701+1.33491&332.7006800000&1.3349100000	&&19.68&19.94&19.61&19.63&19.60&O&&&\\			
LAMOST$\_$332.811+2.41766&332.8110400000&2.4176600000&2.1465500000	&18.16&18.00&18.02&17.85&17.68&O&&&\\			
LAMOST$\_$332.309+2.15514&332.3093000000&2.1551400000&1.4237300000&19.17&19.06&18.88&18.84&18.93&O&&&\\			
LAMOST$\_$333.384+2.38959&333.3839700000&2.3895900000&1.2714900000&18.97&18.83&18.55&18.51&18.63&O&&&\\	
  \noalign{\smallskip}\hline
\end{tabular}
\tablecomments{0.86\textwidth}{Only a portion of Table is shown here for illustration. The whole Table contains information of all quasars is available in the online electronic version.}
\end{sidewaystable}

\newpage
\normalem
\begin{acknowledgements}
This work has made use of data from the European Space Agency (ESA) mission Gaia, processed by the Gaia Data Processing and Analysis Consortium (DPAC). The author is greatful to the developers of TOPCAT (\citealt{Taylor2005TOPCAT}) for their software. This publication makes use of data products from the Wide-field Infrared Survey Explorer, which is a joint project of the University of California, Los Angeles, and the Jet Propulsion Laboratory/California Institute of Technology, funded by the National Aeronautics and Space Administration. Guoshoujing Telescope (the Large Sky Area Multi-Object Fiber Spectroscopic Telescope LAMOST) is a National Major Scientific Project built by the Chinese Academy of Sciences. Funding for the project has been provided by the National Development and Reform Commission. LAMOST is operated and managed by the National Astronomical Observatories, Chinese Academy of Sciences. This work has been supported by the grants from the National Science Foundation of China (NSFC) through grants 11703065 and 11573054.

\end{acknowledgements}
  
\bibliographystyle{raa}
\bibliography{bibtex}

\end{document}